\documentclass[twocolumn,superscriptaddress,amsmath,amssymb,aps,longbibliography,prl]{revtex4}
\usepackage[colorlinks=true,urlcolor=blue,citecolor=blue,linkcolor=blue]{hyperref}
\usepackage{txfonts}
\usepackage{amsfonts}
\usepackage{amsmath}
\usepackage{graphicx}
\usepackage{diagbox}
\usepackage{amssymb}
\usepackage{bbm}
\usepackage{float}
\usepackage{subfigure}
\usepackage{url}
\usepackage{hyperref}
\usepackage{multirow}
\usepackage{bm}
\usepackage{booktabs}
\usepackage{array}
\usepackage{color}
\usepackage{longtable}
\usepackage{booktabs}
\usepackage{algorithm}
\usepackage{algorithmic}

\newcommand{\PreserveBackslash}[1]{\let\temp=\\#1\let\\=\temp}
\usepackage{makecell}
\newcolumntype{C}[1]{>{\PreserveBackslash\centering}p{#1}}
\newcolumntype{R}[1]{>{\PreserveBackslash\raggedleft}p{#1}}
\newcolumntype{L}[1]{>{\PreserveBackslash\raggedright}p{#1}}

\usepackage{xcolor}
\definecolor{red}{rgb}{0.7, 0.1, 0.1}
\definecolor{green}{rgb}{0.0, 0.5, 0.0}

\begin{document}

\title{Neural Canonical Transformations for Quantum Anharmonic Solids of Lithium}

\author{Qi Zhang}
\affiliation{Beijing National Laboratory for Condensed Matter Physics and Institute of Physics, Chinese Academy of Sciences, Beijing 100190, China}

\author{Xiaoyang Wang}
\affiliation{Laboratory of Computational Physics, Institute of Applied Physics and Computational Mathematics, Fenghao East Road 2, 100094 Beijing, China}

\author{Rong Shi}
\affiliation{CAS Key Laboratory of Quantum Information,
University of Science and Technology of China, Hefei 230026, Anhui, China}
\affiliation{Beijing National Laboratory for Condensed Matter Physics and Institute of Physics, 
Chinese Academy of Sciences, Beijing 100190, China}

\author{Xinguo Ren}
\email{renxg@iphy.ac.cn}
\affiliation{Beijing National Laboratory for Condensed Matter Physics and Institute of Physics, 
Chinese Academy of Sciences, Beijing 100190, China}

\author{Han Wang}
\email{wang\_han@iapcm.ac.cn}
\affiliation{Laboratory of Computational Physics, Institute of Applied Physics and Computational Mathematics, Fenghao East Road 2, 100094 Beijing, China}
\affiliation{HEDPS, CAPT, College of Engineering, Peking University, 100871 Beijing, China}

\author{Lei Wang}
\email{wanglei@iphy.ac.cn}
\affiliation{Beijing National Laboratory for Condensed Matter Physics and Institute of Physics, 
Chinese Academy of Sciences, Beijing 100190, China}

\date{\today}
	
\begin{abstract}
    Lithium is a typical quantum solid, characterized by cubic structures at ambient pressure.
    As the pressure increases, it forms more complex structures and undergoes a metal-to-semiconductor transformation, complicating theoretical and experimental analyses.
    We employ the neural canonical transformation approach, an \textit{ab initio} variational method based on probabilistic generative models, to investigate the quantum anharmonic effects in lithium solids at finite temperatures.
    This approach combines a normalizing flow for phonon excited-state wave functions with a probabilistic model for the occupation of energy levels, optimized jointly to minimize the free energy.
    Our results indicate that quantum anharmonicity lowers the \textit{bcc}-\textit{fcc} transition temperature compared to classical molecular dynamics predictions.
    At high pressures, the predicted fractional coordinates of lithium atoms in the \textit{cI16} structure show good quantitative agreement with experimental observations.  
    Finally, contrary to previous beliefs, we find that the poor metallic \textit{oC88} structure is stabilized by the potential energy surface obtained via high-accuracy electronic structure calculations, rather than thermal or quantum effects.
\end{abstract}
\maketitle

\textit{Introduction}.--
Accurate prediction of crystal structures has long been a central focus in materials science.
At low temperatures, a deep understanding of the physical properties of crystals composed of light elements typically requires proper treatment of nuclear quantum effects with anharmonicity~\cite{cazorla2017simulation, monserrat2013anharmonic, kapil2019assessment, flores2020perspective, monacelli2021sscha}.
These effects can play a crucial role in determining the crystal structure,
as seen in hydrogen~\cite{azadi2014dissociation, borinaga2016anharmonic, monacelli2021black, monacelli2023quantum}, helium~\cite{pederiva1998does, vitali2010ab, monserrat2014electron}, and hydrides~\cite{ion2013first, ion2014anharmonic, errea2016quantum, errea2020quantum}.
In this study, we explore one of the most notable examples: lithium, the lightest alkali metal, where the quantum effects of nuclei are pronounced in a wide range of pressures and temperatures~\cite{guillaume2011cold}.
Although lithium behaves as a nearly free-electron metal at low pressure and adopts simple, high-symmetry cubic structures,
the free energy difference between its \textit{bcc} (body-centered cubic) and \textit{fcc} (face-centered cubic) structures is less than $1~\text{meV/atom}$~\cite{hutcheon2019structure, jerabek2022solving, riffe2024alkali, phuthi2024vibrational}, making precise calculations challenging. 
Additionally, lithium exhibits several metastable structures that further complicate experimental measurements~\cite{ackland2017Quantum}.
As the pressure increases, lithium exhibits complex physical behaviors, 
such as anomalous melting curves~\cite{guillaume2011cold, schaeffer2012high, frost2019high}, 
and intriguing phase transitions from metal to semiconductor and back~\cite{matsuoka2009direct, lv2011predicted, marques2011crystal}. 
Moreover, some high-pressure phases consist of large unit cells with tens or even hundreds of atoms~\cite{marques2011crystal, gorelli2012lattice, wang2023data}, posing substantial challenges for both theoretical and experimental studies.

Numerical approaches to studying quantum crystals at finite temperatures include the well-established path integral molecular dynamics~\cite{martyna1999molecular} and path integral Monte Carlo~\cite{barker1979a}. 
In recent years, inspired by the successful application of vibrational self-consistent field theory in molecular studies~\cite{bowman1978self, bowman1986self, gerber1979semiclassical}, efforts have been made to extend it to study crystals~\cite{monserrat2013anharmonic, monserrat2014electron, hutcheon2019structure, kapil2019assessment}. 
However, it relies on a Taylor expansion of the Born-Oppenheimer energy surface (BOES), and the wave function is a simple Hartree product.
The stochastic self-consistent harmonic approximation (SSCHA)~\cite{ion2013first, ion2014anharmonic, monacelli2021sscha} provides an alternative by accounting for both ionic quantum and anharmonic effects without assumptions on the specific function form of the BOES. 
Nevertheless, it still relies on the Gaussian variational density matrices to define the quantum probability distribution. 
Recent developments have extended SSCHA to non-Gaussian assumptions, yet the entropy is still restricted to Gaussian approximations~\cite{siciliano2024beyond, monacelli2024unified}.

In this work, we utilize the recently developed neural canonical transformations (NCT) approach~\cite{xie2022abinitio, xie2023mstar,zhang2024neural}, which is an \textit{ab initio} variational density matrix method based on deep generative models, to study quantum lattice dynamics of lithium. 
NCT constructs orthogonal variational wave functions to describe phonons through a normalizing flow model~\cite{dinh2014nice, dinh2016density, wang2018generatemodels, papamakarios2019neural, papamakarios2021normalizing, xie2022abinitio, xie2023mstar, saleh2023computing, zhang2024neural}.
Additionally, we create a probabilistic model to describe the classical energy occupation probabilities for these phonons, allowing for accurate entropy calculations.
For electronic calculations, we employ the deep potential model~\cite{wang2018deepmd, zeng2023deepmd, wang2023data}, a machine learning BOES, offering significant computational efficiency improvements over density functional theory (DFT) calculations.  
NCT's key advantage is its ability to integrate quantum and anharmonic effects of nuclei into the wave functions, which facilitates the determination of the phonon spectrum. 
Moreover, the independently optimized phonon energy occupation probabilities enable the computation of anharmonic entropy.
The NCT codes for lithium are open-sourced and publicly available~\cite{mycode}.

\textit{The vibrational Hamiltonian of quantum solids}.--
Due to the substantial mass difference between electrons and nuclei, typically spanning several orders of magnitude, the Born-Oppenheimer approximation can be applied to decouple their motions and treat them independently.
The vibrational Hamiltonian is expressed as
$
 H_{\text{vib}} = - \sum_i \frac{1}{2M} \nabla_i^2 + V_{\text{el}}(\bm{R}),
$
where the mass of a lithium atom is $M=6.941~\text{amu}$.
The term $ V_{\text{el}}(\bm{R}) $ is the BOES, derived from electronic structure calculations at nuclear positions $\bm{R}$.
In this work, to ensure both accuracy and computational efficiency, we use the deep potential model to fit the BOES~\cite{wang2018deepmd, zeng2023deepmd, wang2023data}, which is derived from DFT calculations using the Perdew-Burke-Ernzerhof (PBE) functional~\cite{perdew1996generalized}.

The dynamical matrix can be derived from the Hessian of $ V_{\text{el}}$ at the equilibrium position $\bm{R_0}$~\cite{martin2020electronic, baroni2001phonos, souvatzis2008entropy, souvatzis2009the, flores2020perspective, supplemental}:
$
C_{(i\alpha),(j\beta)} 
= \frac{1}{M}
\frac{\mathrm{\partial}^2 V_{\text{el}}}
{\mathrm{\partial} u_{i \alpha} \mathrm{\partial} u_{j \beta}},
$
where $ i, j $ index the nuclei, 
$ \alpha, \beta $ represent the Cartesian components,
and the displacement coordinates are defined as $\bm{u}=\bm{R}-\bm{R_0}$.
Diagonalizing the matrix in a supercell containing $ N $ atoms yields $ D=3N - 3 $ non-zero eigenvalues, which correspond to the number of phonon modes.
The eigenvalues are related to the squares of the phonon frequencies, $ \omega_k^2 $ ~ ($k=1,2,\ldots, D$), 
and the associated eigenvectors define the unitary transformation from displacement coordinates $\bm{u}$ to phonon coordinates $\bm{q}$.
Consequently, the Hamiltonian can be expressed in phonon coordinates:
\begin{equation}
H_{\text{vib}} 
= \frac{1}{2} \sum_{k=1}^D \left(
    - \frac{\partial^2 }{ \partial q_k^2} 
    + \omega_k^2 q_k^2
    \right)
+ V_{\text{anh}}(\bm{q}),
\end{equation}
where the term $V_{\text{anh}}$ represents the anharmonic contributions, as detailed in~\cite{supplemental}.
In this representation, the separation of high and low-frequency modes greatly enhances the efficiency in the following calculations.

\textit{Neural canonical transformation for variational density matrix}.--
The solution for a many-body system in the canonical ensemble can be obtained by minimizing the Helmholtz free energy for a variational density matrix:
\begin{equation}
F = k_B T~\mathrm{Tr}(\rho \ln \rho) + \mathrm{Tr}(\rho H_{\text{vib}}),
\end{equation}
where $k_B$ is the Boltzmann constant and $T$ is the temperature.
Assuming that the phonons occupy the states $| \Psi_{\bm{n}} \rangle$ with probability $p_{\bm{n}}$, the variational density matrix can be represented in terms of these quantum states as:
\begin{equation}
\rho = \sum_{\bm{n}} p_{\bm{n}}~
|       \Psi_{\bm{n}}  \rangle 
\langle \Psi_{\bm{n}}  |,
\label{eq:vdm}
\end{equation}
where $\bm{n}=(n_1, n_2, \ldots, n_D)$ indexes the energy levels of the phonons.
An unbiased estimate of the anharmonic free energy for the variational density matrix Eq.~\ref{eq:vdm} can be written as nested thermal and quantum expectations:
\begin{equation}
F = \underset{\bm{n} \sim p_{\bm{n}}}
    {\mathbb{E}} 
    \left[
    k_B T \ln p_{\bm{n}} + 
    \underset{\bm{q} \sim |\Psi_{\bm{n}}(\bm{q})|^2}
    {\mathbb{E}} \left[
        \frac{H_{\text{vib}} \Psi_{\bm{n}}(\bm{q})}{\Psi_{\bm{n}}(\bm{q})}
    \right]
\right],
\label{eq:helmholtz}
\end{equation}
where $\Psi_{\bm{n}}(\bm{q})=\langle\bm{q}|\Psi_{\bm{n}}\rangle$ is the wave function in phonon coordinates.
The symbol ${\mathbb{E}}$ is the statistical expectation, which can be estimated through sampling~\cite{zhang2024neural, becca2017quantum}.
In this Letter, the variational parameters within the energy occupation probabilities and wave functions are denoted as $\bm{\mu}$ and $\bm{\theta}$, respectively, i.e., $p_{\bm{n}}=p_{\bm{n}}({\bm{\mu}})$, $\Psi_{\bm{n}}(\bm{q})=\Psi_{\bm{n}}(\bm{\theta},\bm{q})$.
These parameters can be optimized via gradient descent~\cite{kingma2014adam}, with $F$ as the loss function.
The gradients $\nabla_{\bm{\mu}}F$ and $\nabla_{\bm{\theta}}F$~\cite{supplemental} can be efficiently computed using automatic differentiation~\cite{jax2018github}.

In a supercell with $D$ vibrational modes, setting a cutoff of $K$ energy levels per phonon (i.e., $n_k=1,2,\ldots, K$) results in an exponentially huge state space of $K^D$. 
For supercells containing hundreds of atoms, directly representing the energy occupation probabilities $p_{\bm{n}}$ becomes impractical in computations.
In the study of lithium,  we assume that the probability distributions take a product form~\cite{martyn2019product}: $
    p_{\bm{n}} = \prod_{k=1}^D p(n_k)$, 
where $p(n_k)$ represents the probability of the $k$-th phonon occupying state $n_k$, 
and they are governed by learnable parameters $\bm{\mu}$. 
We have checked that an even more powerful variational autoregressive network~\cite{wu2019solving, liu2021solving, xie2023mstar} does not improve results, likely due to weak coupling between phonon modes. 
The entropy is the expectation of the probabilities:
\begin{equation}
S = \underset{\bm{n} \sim p_{\bm{n}}}{\mathbb{E}}
    [- k_B \ln p_{\bm{n}}].
\label{eq:entropy}
\end{equation}
We note the nonlinear SSCHA~\cite{siciliano2024beyond, monacelli2024unified} corresponds to even further simplification of $p_{\bm{n}}$, which assumes that the entropy is given by a set of independent harmonic oscillators.

To construct variational wave functions, we apply a unitary transformation to a set of orthogonal basis states~\cite{xie2022abinitio, xie2023mstar, zhang2024neural}:
$|\Psi_{\bm{n}}\rangle = U_{\bm{\theta}}~|\Phi_{\bm{n}}\rangle$,
where the basis states $|\Phi_{\bm{n}}\rangle$ are chosen as the wave functions of a $D$-dimensional harmonic oscillator with frequencies $\omega_k$.
We implement the unitary transformation $U_{\bm{\theta}}$ using a normalizing flow~\cite{dinh2014nice, dinh2016density, wang2018generatemodels, papamakarios2019neural, papamakarios2021normalizing, xie2022abinitio, xie2023mstar, saleh2023computing, zhang2024neural},
which establishes a learnable bijection between the phonon coordinates $\bm{q}$ and a set of quasi-phonon coordinates $\bm{\xi}$.
The bijection is represented as a smooth, reversible function $\bm{\xi}=f_{\bm{\theta}}(\bm{q})$, where $f_{\bm{\theta}}$ consists of neural networks with learnable parameters $\bm{\theta}$,
specifically, a real-valued non-volume preserving network~\cite{dinh2016density}.
Accordingly, the orthogonal variational wave functions of all energy levels can be formulated as~\cite{supplemental, zhang2024neural}:
\begin{equation}
    \Psi_{\bm{n}}(\bm{q}) 
    = \Phi_{\bm{n}}\left(f_{\bm{\theta}}(\bm{q})\right) \left| \mathrm{det} 
    \left(
    \frac{\partial f_{\bm{\theta}}(\bm{q})}{\partial \bm{q}} 
    \right)
    \right|^{1/2},
    \label{eq:flow}
\end{equation}
where $\Phi_{\bm{n}}(\bm{\xi}) = \langle \bm{\xi}| \Phi_{\bm{n}} \rangle$ are basis states.
Notably, in the study of lithium, the computation involves about ten million orthogonal states for each training.
The Jacobian determinant in Eq.~\ref{eq:flow} captures phonon interactions and anharmonic effects, enabling a more flexible and accurate representation.
This form outperforms the Gaussian-type assumption in SSCHA, yielding significantly better energy and quantum distributions in an anharmonic potential benchmark, as detailed in Supplemental Material (SM)~\cite{supplemental}.
NCT remains robust when imaginary phonons appear in strong anharmonicity systems (e.g., saddle points of BOES). 
In such cases, we can choose the corresponding basis states with real-valued frequencies, and the flow model will automatically optimize to find the most suitable wave functions.
A detailed derivation of NCT can be found in SM~\cite{supplemental} and 
our previous work~\cite{zhang2024neural}.

We can extend NCT naturally to the isothermal-isobaric ensemble, 
where the goal is to minimize the Gibbs free energy at a target pressure $P^*$, defined as:
\begin{equation}
    G = F + P^*\Omega,
    \label{eq:gibbs}
\end{equation}
where $\Omega$ is the system volume.
From the relation $dG = dF + \Omega \sum \sigma_{\alpha\beta} ~d\varepsilon_{\alpha\beta} $, once the parameters $\bm{\mu}$ and $\bm{\theta}$ have converged under constant volume optimization (i.e., when $dF = 0$), the gradient of the Gibbs free energy to the strain $\bm{\varepsilon}$ depends only on the stress tensor $\bm{\sigma}$.
The stress tensor and pressure can be calculated using the virial theorem~\cite{nielsen1985quantum, martin2020electronic, supplemental}. 
Then, we can optimize the lattice constants $\bm{a}$ through the strain tensor $\varepsilon_{\alpha\beta} = \Omega (\sigma_{\alpha\beta} - P^* \delta_{\alpha\beta})$, which is similar to the structure relaxation in other methods~\cite{monserrat2013anharmonic, monacelli2021sscha}.

\begin{figure}[htbp]
	\centering
    \hspace{-3.7mm}
	\includegraphics[width=0.50\textwidth]{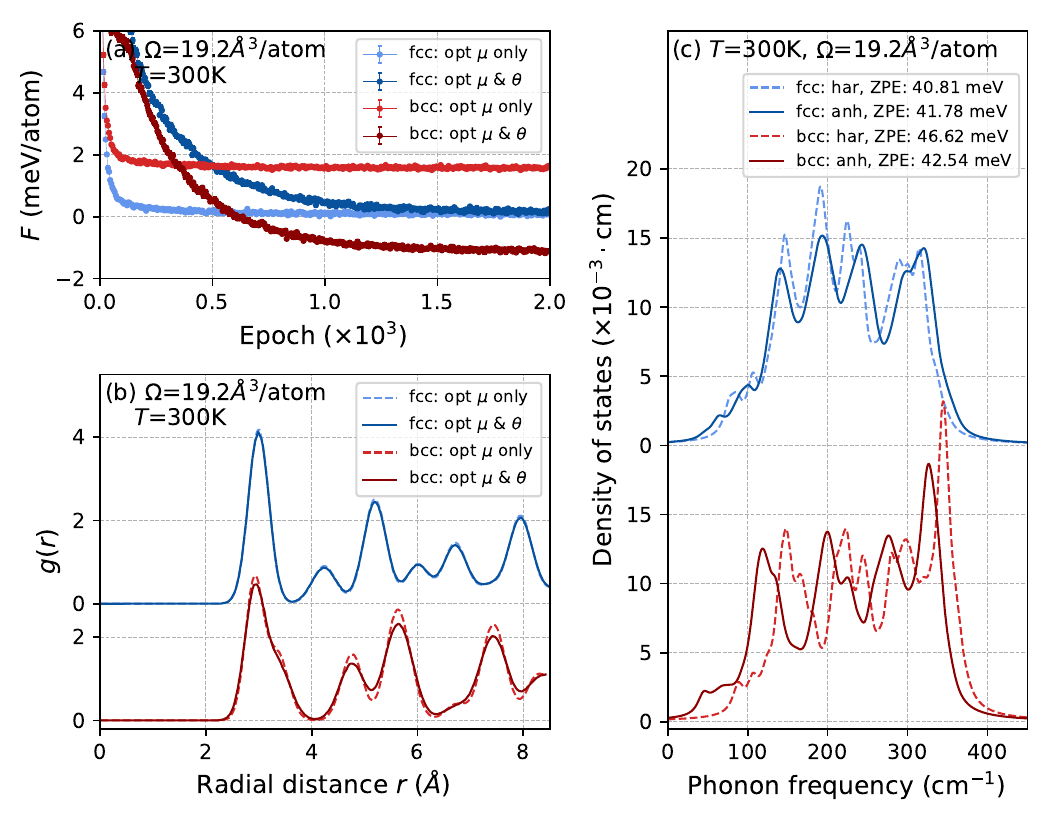}
	\caption{
    Numerical results for \textit{fcc} and \textit{bcc} at a fixed volume of $\Omega = 19.2 ~ \AA^3/\text{atom}$ and temperature $T = 300 ~ \text{K}$.     
    (a) Training curves of the Helmholtz free energy $F(\bm{\mu},\bm{\theta})$ (Eq.~\ref{eq:helmholtz}).
    The legend ``opt $\bm{\mu}$ only" indicates that only the energy occupation probabilities $p_{\bm{n}}$ are optimized, and ``opt $\bm{\mu}$ \& $\bm{\theta}$" means that both $p_{\bm{n}}$ and $\Psi_{\bm{n}}$ are optimized.
    (b) Radial distribution functions of nuclei.
    (c) Phonon density of states per atom. 
    The harmonic (har) frequencies $\omega_k$ are calculated from the dynamical matrix, and the anharmonic (anh) frequencies $w_k$ are taken from the single-phonon excitations.
    The zero-point energies (ZPE) are defined as $E_{\text{ZPE}, \text{har}} = \sum_{k=1}^D \omega_k/{2N}$ and $E_{\text{ZPE}, \text{anh}} = \sum_{k=1}^D w_k/{2N}$.
    }
	\label{fig:1}
\end{figure}

\textit{Anharmonic and nuclear quantum effects on stability}.--
At ambient conditions, lithium adopts a simple \textit{bcc} structure. 
As the temperature decreases, experiments have demonstrated that the true ground state of lithium is \textit{fcc}~\cite{ackland2017Quantum}. 
Some calculations have revealed that the free energies of these structures are extremely close~\cite{hutcheon2019structure, jerabek2022solving, riffe2024alkali, phuthi2024vibrational}, highlighting the necessity of fully accounting for quantum and anharmonic effects.
To investigate the influence of anharmonicity, 
we first conducted NCT calculations for \textit{bcc} and \textit{fcc} using supercells with $250$ and $256$ atoms, respectively, 
at a fixed volume of $ 19.2 ~ \AA^3/\text{atom}$ and temperature $ 300 ~ \text{K}$.
As a comparative study, we set $f_{\bm{\theta}}$ in Eq.~\ref{eq:flow} to an identity transformation, meaning the phonon wave functions are harmonic oscillators.
In this case, only the phonon occupation probabilities $p_{\bm{n}}$ in Eq.~\ref{eq:helmholtz} were optimized.

At a lower temperature of $100~\text{K}$, the free energies of \textit{fcc} are lower than that of \textit{bcc}, as expected. 
However, as the temperature increases to $300~\text{K}$, the impact of anharmonicity becomes evident, as shown in FIG.~\ref{fig:1}~(a). 
In \textit{fcc}, the free energy difference between the two approaches remains small, about $0.11 ~ \text{meV/atom}$. 
In contrast, the anharmonic effects are much stronger in \textit{bcc}, and the difference expands to $2.67 ~ \text{meV/atom}$.
It is also observed that when we only optimized $p_{\bm{n}}$, the free energy of \textit{fcc} is lower than that of \textit{bcc}. 
However, when the optimization of wave functions is included, i.e., when anharmonic effects are considered, the \textit{bcc} structure becomes more stable.
This phenomenon suggests that \textit{bcc} exhibits stronger anharmonicity than \textit{fcc}, underscoring the critical role of anharmonicity in determining the stability.

The findings are further supported by the radial distribution functions (RDF) of nuclei, as shown in FIG.~\ref{fig:1}~(b). 
The RDF for \textit{fcc} exhibits only slight influence from anharmonic effects.
In contrast, the RDF for \textit{bcc} exhibits a smoother curve when anharmonic effects are considered, indicating a reduction in atomic localization.
This behavior suggests that anharmonicity softens the system, resulting in a lower zero-point energy (ZPE) than the harmonic approximation.
Further insights are provided by the phonon density of states (DOS) depicted in FIG.~\ref{fig:1}~(c), where the ZPE is determined as half the sum of all phonon frequencies per atom, as detailed in SM~\cite{supplemental}.
Although the \textit{bcc} structure is more stable at high temperatures, the numerical results reveal that the ZPE of \textit{fcc} remains lower than that of \textit{bcc}.
This phenomenon can be explained by the differences in coordination numbers: 
the coordination number of \textit{bcc} is $8$, while that of \textit{fcc} is $12$. 
Hence, atoms in \textit{bcc} interact less strongly with their neighbors, 
leading to higher quantum fluctuations and greater anharmonicity.


\begin{figure}[htbp]
	\centering
    \hspace{-3.7mm}
	\includegraphics[width=0.50\textwidth]{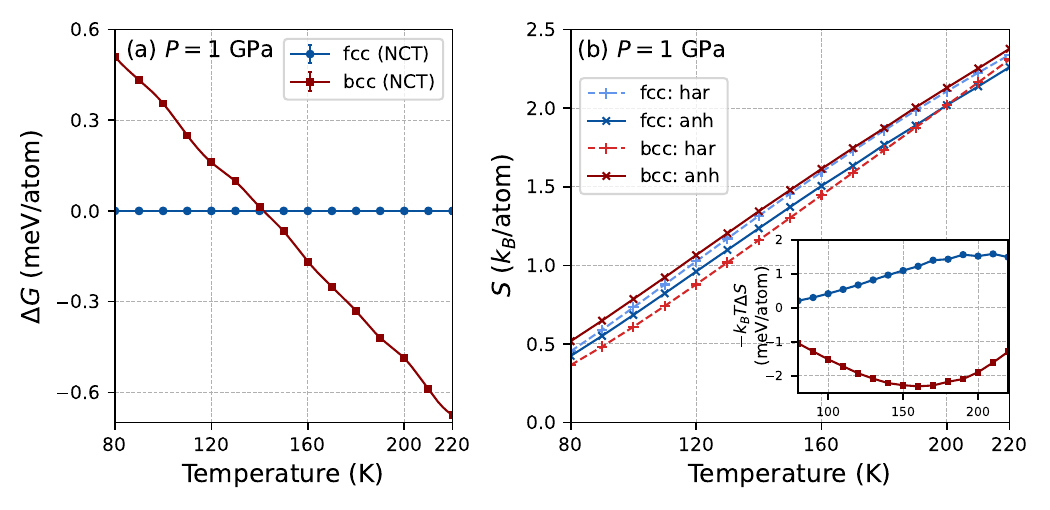}
	\caption{
    (a) Gibbs free energy (Eq.~\ref{eq:gibbs}) difference between \textit{fcc} and \textit{bcc} at $P=1~\text{GPa}$, using \textit{fcc} as the reference.
    The error bars represent statistical uncertainties, which are smaller than the data points. For additional sources of uncertainty, refer to~\cite{supplemental}.
    (b) Anharmonic effects on ionic entropy.
    The anharmonic (anh) entropy is directly obtained from the expectation of the probability distribution (Eq.~\ref{eq:entropy}), while the harmonic (har) entropy is calculated from the harmonic frequencies using $S = \sum_k \left[ \frac{{\omega_k}}{k_B T} \frac{1}{\text{e}^{ {\omega_k} /k_B  T} - 1} - \ln (1 - \text{e}^{-{\omega_k} /k_B  T})\right]$. 
    The $x$-axis of the inset represents the temperature, and the $y$-axis corresponds to $- k_B T (S_\text{anh} - S_\text{har})$ in units of $\text{meV/atom}$. 
    }
	\label{fig:2}
\end{figure}

To gain deeper insight into the influence of nuclear quantum effects, we performed calculations on the \textit{bcc} and \textit{fcc} structures under constant pressure.
The Gibbs free energies of both structures are extremely close~\cite{hutcheon2019structure, jerabek2022solving, riffe2024alkali, phuthi2024vibrational}. 
An error of just $1~\text{meV}$ could lead to a shift of more than $100~\text{K}$ in the transition temperature~\cite{riffe2024alkali}. 
FIG.~\ref{fig:2}~(a) shows the Gibbs free energy difference between these structures at $1~\text{GPa}$, with the \textit{fcc} structure used as the reference. 
The two curves intersect at $142~\text{K}$, indicating a phase transition at this temperature.
We also calculated the transition temperature through classical molecular dynamics (MD) simulations with thermodynamic integration on the same BOES, obtaining a value of $185~\text{K}$.
The main difference between these methods is that NCT accounts for the quantum effects, while MD simulations do not.
Similar results are also observed at $0$ and $2~\text{GPa}$, as detailed in SM~\cite{supplemental}, where NCT consistently predicts lower transition temperatures compared to MD.

The ionic entropy of both structures is shown in FIG.~\ref{fig:2}~(b). 
The anharmonic entropy obtained from NCT is derived directly from the probabilities of energy occupations in Eq.~\ref{eq:entropy}, beyond the harmonic oscillator assumption.
The entropy of \textit{fcc} is higher than that of the \textit{bcc} under the harmonic oscillator assumption.
However, when the anharmonicity is considered, the relationship is reversed.
The higher entropy of the \textit{bcc} structure is a key factor contributing to its stability in finite temperatures~\cite{souvatzis2008entropy, souvatzis2009the, soderlind2012high, hutcheon2019structure, phuthi2024vibrational}.
Furthermore, we quantified the free energy difference arising from the anharmonic effects of entropy as $-k_B T~(S_{\text{anh}} - S_{\text{har}})$ (inset of FIG.~\ref{fig:2}~(b)), estimating it to be on the order of several $\text{meV}$.
This underscores the critical importance of accurately incorporating anharmonic effect in the calculations.


\begin{figure}[htbp]
	\centering
    \hspace{-3.7mm}
	\includegraphics[width=0.50\textwidth]{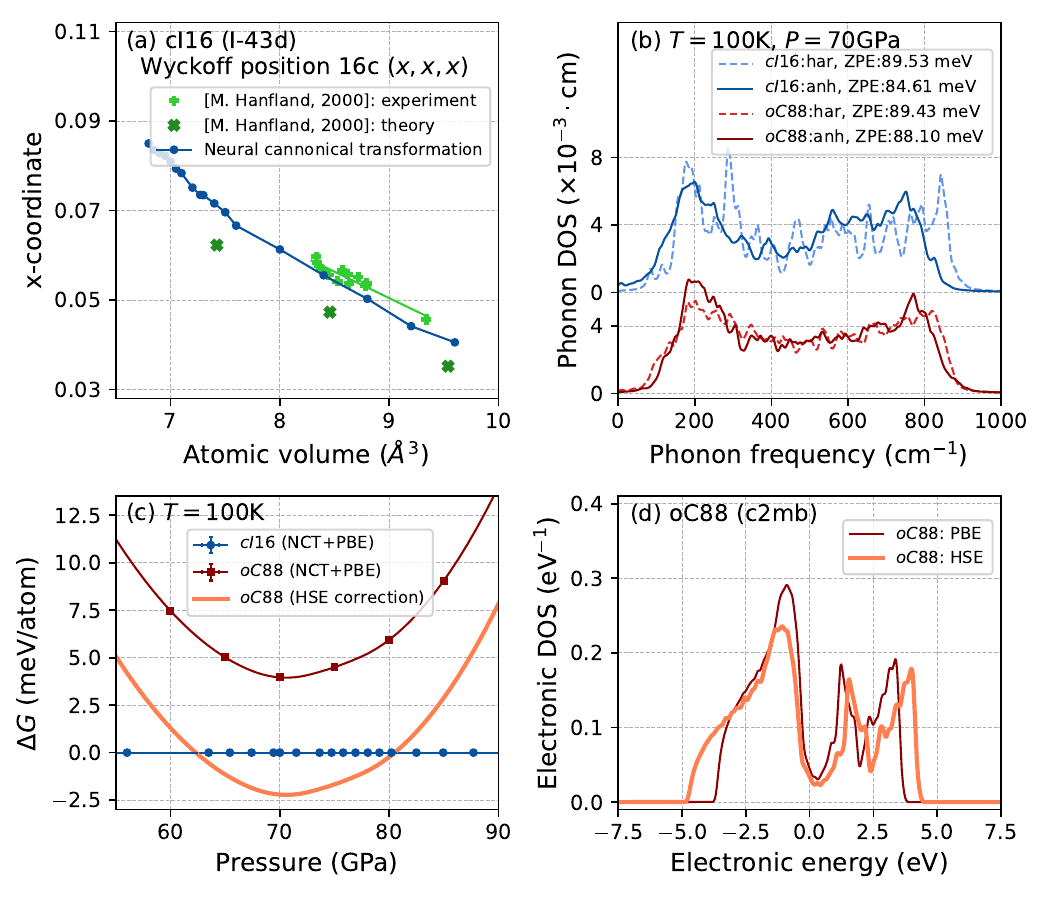}
	\caption{
    (a) Fractional coordinates of the Wyckoff position $16c$ in the \textit{cI16} (\textit{I-43d}) structure.
    Experimental values are taken from Ref.~\cite{hanfland2000new}.
    (b) Phonon density of states (DOS). 
    (c) Gibbs free energy difference between \textit{cI16} and \textit{oC88} at $T=100~\text{K}$, with \textit{cI16} taken as the reference.
    The thick orange line indicates the single-point correction for \textit{cI16} and \textit{oC88} at $70~\text{GPa}$, based on higher-accuracy electronic structure calculations using the HSE functional, as compared to the PBE. 
    (d) Electronic DOS for \textit{oC88} with PBE and HSE functionals.
    } 
	\label{fig:3}
\end{figure}

\textit{High-pressure structural stability of lithium}.--
Under high pressure, lithium exhibits more complex structures and larger unit cell sizes.
We first applied the NCT method to calculate the \textit{cI16} (cubic I-centered, \textit{I-43d}) structure at $100~\text{K}$, using a supercell of $432$ atoms under various pressures. 
The NCT method optimizes the atom positions through coordinate transformations.
Subsequently, we quenched the sampled structures to their ground state with the BOES and analyzed the fractional coordinates $(x, x, x)$ of Wyckoff position $16c$ as a function of atomic volume.
As depicted in FIG.~\ref{fig:3}~(a), our results are consistent with the experiment reported in Ref.~\cite{hanfland2000new}, demonstrating the reliability of NCT in structure optimizations.

As illustrated in FIG.~\ref{fig:3}~(b), the phonon DOS and ZPE of the \textit{cI16} and \textit{oC88} (orthorhombic C-face centered, \textit{C2mb}) structures are calculated at $70~\text{GPa}$.
Under the harmonic approximation, the ZPE of both structures are found to be comparable, consistent with the results obtained using the finite displacement and density functional perturbation theory methods~\cite{wang2023data}.
The anharmonic effects soften the phonon spectrum of \textit{cI16}, reducing the ZPE by $4.92~\text{meV}$ and further enhancing its stability.
In contrast, the ZPE of \textit{oC88} decreases by only $1.67~\text{meV}$ under anharmonic effects, indicating a smaller impact compared to \textit{cI16}.
This result suggests that when the anharmonic effect is considered, the stability of \textit{oC88} decreases, contrary to the expectations of previous studies~\cite{marques2011crystal,lv2011predicted, wang2023data}.

Additionally, we calculated the Gibbs free energies at $100~\text{K}$, as shown in FIG.~\ref{fig:3}~(c).
The free energy of \textit{oC88} remains consistently higher than that of \textit{cI16} across all pressures, which contradicts previous experiments.
It has been reported that the resistivity increases sharply by more than four orders of magnitude after the \textit{cI16} phase, ultimately transforming into a semiconductor.
Compression experiments in Ref.~\cite{guillaume2011cold} observed the \textit{cI16}-\textit{oC88} transition, as evidenced by changes in crystal color and diffraction patterns. 
Raman spectra measurements in Ref.~\cite{gorelli2012lattice} detected signals corresponding to the \textit{oC88} phase. 
Another experiment~\cite{frost2019high} also observed a phase transition around $60~\text{GPa}$ through the changes in diffraction peaks.

After the \textit{oC88} structure was experimentally observed~\cite{guillaume2011cold}, theoretical studies attempted to explain its stability.
However, Ref.~\cite{marques2011crystal} concluded from their zero-temperature calculations that \textit{oC88} was only the second most stable phase with an enthalpy about $1~\text{meV}$ higher than \textit{cI16}, attributing this to an insufficient consideration of ZPE and thermal effects.
In another work~\cite{lv2011predicted}, the authors also failed to identify \textit{oC88} as a stable structure.
In contrast, Ref.~\cite{gorelli2012lattice} found that \textit{oC88} is stable when using the PBE functional with a harmonic ZPE at $200~\text{K}$.
However, a recent study~\cite{wang2023data} demonstrated that neither harmonic nor anharmonic approximations could reproduce the results of Ref.~\cite{gorelli2012lattice} at various conditions, showing a free energy difference at least $1~\text{meV}$ with \textit{oC88} consistently higher than \textit{cI16}.
Surprisingly, as NCT captures nuclear quantum effects and anharmonic behaviors more accurately, the difference increases to $4~\text{meV}$.
It has been observed that DFT often over-stabilizes metallic states relative to non-metallic states~\cite{ma2009transparent, ma2007structure}.
Therefore, we strongly suspect that the instability of \textit{oC88} arises from the limited accuracy of the DFT (PBE) calculations used in fitting the BOES~\cite{wang2023data}.

To validate our hypothesis, we employed the NCT-optimized structures at $70~\text{GPa}$ and conducted single-point electronic structure calculations using the high-accuracy Heyd-Scuseria-Ernzerhof (HSE06) functional~\cite{martin2020electronic}.
The HSE functional incorporates a hybrid exchange-correlation correction, enabling a clearer distinction between metallic and non-metallic states.
Notably, HSE calculations are significantly more computationally demanding, requiring approximately two orders of magnitude more computational resources than PBE. 
Additional details of HSE are provided in SM~\cite{supplemental}.
Our results reveal that the relative energy of \textit{oC88} compared to \textit{cI16} decreases by $6.17~\text{meV}$ under HSE in comparison to PBE.
This reduction is significantly larger than the contributions from ZPE, anharmonic, and finite temperature effects.
The HSE correction, depicted as the thick line in FIG.~\ref{fig:3}~(c), predicts a \textit{cI16}-\textit{oC88} phase transition at approximately $62~\text{GPa}$ and $100~\text{K}$. 
This finding is consistent with experimental observations, which report a narrow stability range for the \textit{oC88} phase, existing between $62$ and $70~\text{GPa}$, flanked by the \textit{cI16} and \textit{oC40} phases on either side, respectively~\cite{guillaume2011cold}.
The electronic DOS of \textit{oC88}, depicted in FIG.~\ref{fig:3}~(d), shows that while the HSE correction lowers the potential energy, \textit{oC88} still behaves as a poor metal.

\textit{Conclusions}.--
In summary, we developed the NCT method~\cite{xie2022abinitio, xie2023mstar,zhang2024neural} to study anharmonic quantum solids and applied it to lithium. 
It enables the calculation of excited-state wave functions of nuclear motions beyond the harmonic approximation, allowing for the extraction of anharmonic phonon spectra. 
The independently optimized phonon occupation probabilities facilitate the computation of anharmonic entropy. 
The results demonstrate that quantum anharmonic effects play a crucial role in structural stability and introduce significant corrections to the \textit{fcc}-\textit{bcc} transition temperature. 
The fractional coordinates of the \textit{cI16} structure have been determined and closely align with experimental findings.
Moreover, we identified that the failure of previous numerical studies~\cite{marques2011crystal,lv2011predicted, wang2023data} to observe the stability of \textit{oC88} was due to the limitations of the PBE functional in accurately describing poor metallic states.
To address this, we applied the HSE functional to refine the results and estimate the stability region of \textit{oC88}. 
Looking ahead, both experimental and computational investigations suggest that the emergence of novel high-density lithium solid structures between the \textit{cI16} and liquid phases presents a promising avenue for future exploration~\cite{frost2019high, wang2023data}.
Overall, NCT shows significant potential for investigating other light-element systems, such as hydrogen~\cite{azadi2014dissociation, borinaga2016anharmonic, monacelli2021black, monacelli2023quantum}, helium~\cite{pederiva1998does, vitali2010ab, monserrat2014electron}, and hydride solids~\cite{ion2013first, ion2014anharmonic, errea2016quantum, errea2020quantum}, as well as molecular systems like aspirin and paracetamol~\cite{sauceda2021dynamical}, where quantum anharmonicity plays a crucial role.
Similar to the techniques used in SSCHA~\cite{ion2013first}, NCT could also be extended to calculate electron-phonon coupling, which is important for studying superconductivity.
It could greatly enhance our understanding and address a wide range of challenges in quantum crystals.

\textit{Acknowledgements}.--
We are grateful for the valuable discussions with Hao Xie, Jian Lv, Xinyang Dong, Zhendong Cao, Zihang Li, Ruisi Wang, and Peize Lin.
Some calculations have been done on the supercomputing system in the Huairou Materials Genome Platform.
This work is supported by the National Natural Science Foundation of China under Grants No. 92270107, No. T2225018, No. 12188101, No. T2121001, No. 12134012, and No. 12374067,
and the Strategic Priority Research Program of the Chinese Academy of Sciences under Grants No. XDB0500000 and No. XDB30000000.

\bibliographystyle{apsrev4-1}
\bibliography{mybibtex}

\clearpage

\appendix

\begin{widetext}

\begin{center}
    {\large \textbf{Supplemental Material: Neural Canonical Transformations for Quantum Anharmonic Solids of Lithium}}
\end{center}

\setcounter{table}{0}
\setcounter{figure}{0}
\setcounter{equation}{0}
\setcounter{section}{0}
\renewcommand{\theequation}{S\arabic{equation}}
\renewcommand{\thefigure}{S\arabic{figure}}
\renewcommand{\thetable}{S\arabic{table}}
\renewcommand{\thesection}{S\Roman{section}}

\tableofcontents

\section{SI. Neural Canonical Transformation approach}


\begin{figure*}[htbp]
	\centering
	\includegraphics[width=0.85\textwidth]{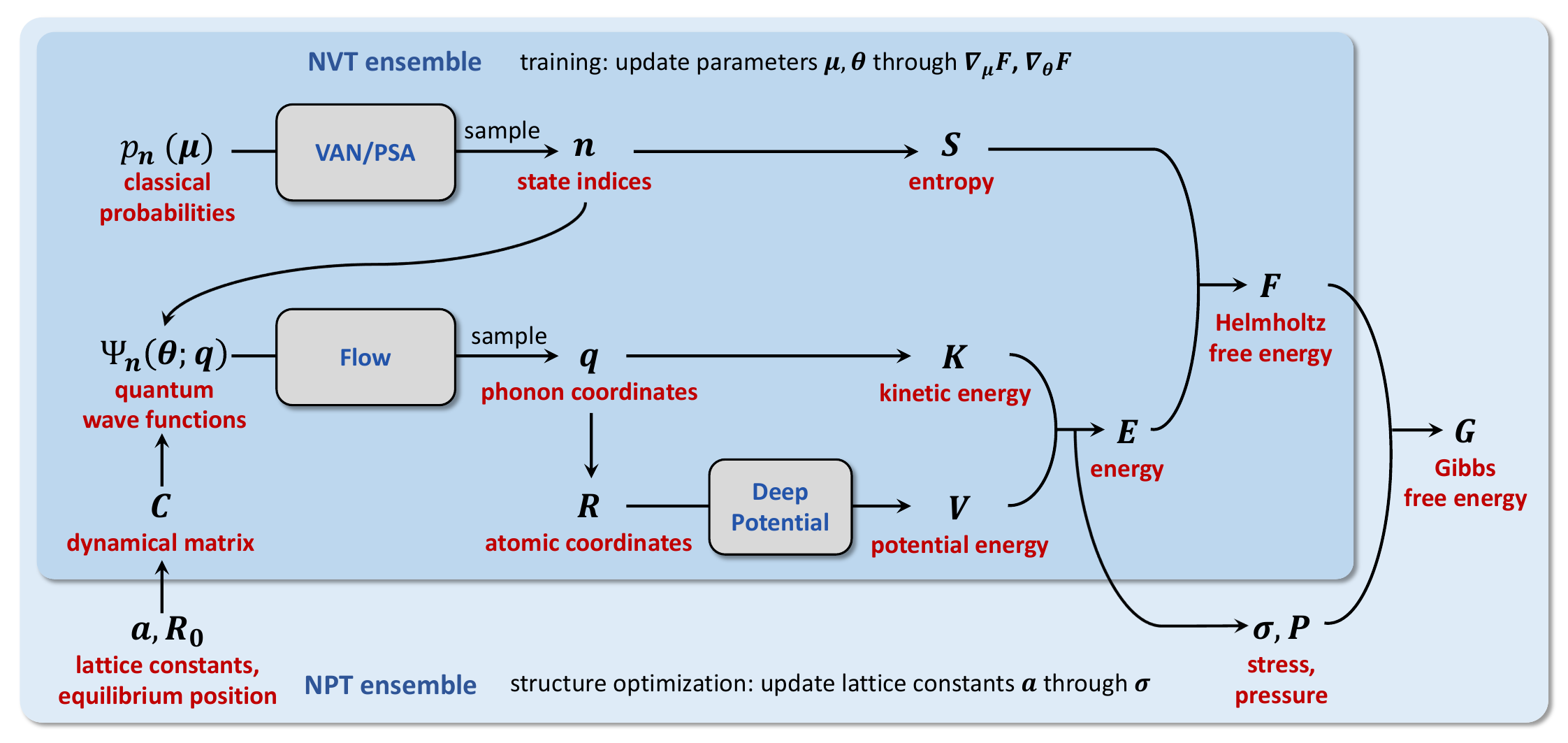}
	\caption{A sketch of the workflow for the neural canonical transformations. 
    }
	\label{Fig:1}
\end{figure*}

Neural canonical transformation (NCT) is a variational density matrix method that leverages multiple deep generative models.
As shown in FIG.~\ref{Fig:1}, 
this approach utilizes either a product spectrum ansatz (PSA)~\cite{martyn2019product} or a variational autoregressive network (VAN)~\cite{wu2019solving, liu2021solving, xie2023mstar} to represent phonon energy occupation probabilities (phonon Boltzmann distribution), a normalizing flow~\cite{dinh2014nice, dinh2016density, wang2018generatemodels, papamakarios2019neural, papamakarios2021normalizing, xie2022abinitio, xie2023mstar, saleh2023computing, zhang2024neural} for phonon excited-state wave functions, and a deep potential (DP) model~\cite{wang2018deepmd, zeng2023deepmd, wang2023data} to describe the Born-Oppenheimer energy surface (BOES).
In constant-volume calculations, the goal is to minimize the Helmholtz free energy $F$ by jointly optimizing the wave functions and probability distributions.
For constant-pressure calculations, once $F$ has converged, the stress tensor $\bm{\sigma}$ is measured to adjust the lattice constants $\bm{a}$. 
After making these adjustments, constant-volume calculations are repeated until the Gibbs free energy $G$ converges.
All converged results of NCT correspond to thermodynamic equilibrium states, where the system's macroscopic properties remain stable over time, with no net energy exchange or unbalanced forces driving further evolution.

\subsection{A. Variational wave functions in phonon coordinates}

The displacement coordinates in the crystal are defined as $\bm{u} = \bm{R} - \bm{R_0}$, 
where $\bm{R}$ represents the nuclear positions and $\bm{R_0}$ denotes the initial equilibrium positions.
The equilibrium geometry is given by the absence of net forces on the nuclei.
The BOES is expressed as a function of the coordinates and the lattice constants: 
$V_{\text{el}}(\bm{a}, \bm{R}) = V_{\text{el}}(\bm{a}, \bm{R_0} + \bm{u})$, 
where we omit the lattice constants $\bm{a}$ in $V_{\text{el}}$ in the main text for simplicity.
The dynamical matrix of the lattice can then be derived from the following expression~\cite{monserrat2013anharmonic, martin2020electronic, baroni2001phonos, souvatzis2008entropy, souvatzis2009the, flores2020perspective}:
\begin{equation}
C_{(i\alpha),(j\beta)} 
= 
\left.
\frac{1}{M}
\frac{\mathrm{\partial}^2 V_{\text{el}}}
{\mathrm{\partial} u_{i \alpha} \mathrm{\partial} u_{j \beta}}
\right|_{\bm{u}=\bm{0}}
,
\label{eq:dynamicalmatrix}
\end{equation}
where $ i, j $ index the nuclei, 
$ \alpha, \beta $ represent the Cartesian components,
and $M$ is the mass of atoms.
By diagonalizing the dynamical matrix, the system can be expressed in phonon coordinates.
The eigenvalues correspond to the squares of phonon frequencies, $ \omega_k^2 $, 
where $k$ indices the phonon modes.
The associated eigenvectors define the unitary transformation $\bm{P}$ between displacement and phonon coordinates:
$\bm{q} = {\sqrt{M}} \bm{u} \bm{P}$, 
$\bm{u} = \frac{1}{\sqrt{M}} \bm{P}^{\dagger} \bm{q}$.
Consequently, the anharmonic terms in the BOES can be expressed as:
\begin{equation}
 V_{\text{anh}}(\bm{q}) = 
 V_{\text{el}}(\bm{a},~\bm{R_0} + \frac{1}{\sqrt{M}} \bm{P}^{\dagger} \bm{q} ) - \sum_k \frac{1}{2} \omega_k^2 q_k^2.
\end{equation}
The vibrational Hamiltonian of the system, including anharmonic terms, can then be expressed in phonon coordinates as:
\begin{equation}
    H_{\text{vib}} 
    = \frac{1}{2} \sum_{k=1}^D \left(
        - \frac{\partial^2 }{ \partial q_k^2} 
        + \omega_k^2 q_k^2
        \right)
    + V_{\text{anh}}(\bm{q}),
    \label{eq:Hvib}
\end{equation}
where the atomic mass $M$ is absorbed in the transformation from displacement coordinates $\bm{u}$ to phonon coordinates $\bm{q}$.
In the harmonic approximation, we discard the anharmonic terms, and the system's Hamiltonian is defined in quasi-phonon coordinates:
\begin{equation}
    H_{\text{basis}} 
    = \frac{1}{2} \sum_{k=1}^D \left(
        - \frac{\partial^2 }{ \partial \xi_k^2} 
        + \omega_k^2 \xi_k^2
    \right).
    \label{eq:Hbasis}
\end{equation}
The eigenvectors of the above Hamiltonian are product states of $D$-dimensional quantum harmonic oscillators,
which serve as the basis of variational wave functions mentioned in the main text:
\begin{equation}
    \Phi_{\bm{n}}(\bm{\xi}) = \prod_{k=1}^D \phi_{n_k}(\xi_k),
    ~~\text{with}~~
    \phi_{n_k}(\xi_k) = e^{-\frac{1}{2} \omega_k \xi_k^2} h_{n_k}(-\sqrt{\omega_k} \xi_k),
\end{equation}
where the normalization factor has been omitted, as only derivatives of the wave function's logarithm are required for subsequent calculations.
$n_k$ is the energy level of the $k$-th quasi-phonon, and $h_{n_k}$ is the $n_k$-th order Hermite polynomial.
The variational wave functions in phonon coordinates $\bm{q}$ are expressed as~\cite{zhang2024neural}:
\begin{equation}\label{eq:wf}
    \Psi_{\bm{n}}(q_1, q_2, \cdots, q_D) 
    = \phi_{n_1}(\xi_1) \phi_{n_2}(\xi_2) \cdots \phi_{n_D}(\xi_D)
    \sqrt{
    \left| 
    \mathrm{det} 
    \left(
    \frac{\partial (\xi_1, \xi_2, \cdots, \xi_D)}
    {\partial (q_1, q_2, \cdots, q_D)} 
    \right)
    \right|
    },
\end{equation}
where the orthogonality condition is always satisfied due to the properties of the Jacobian determinant, i.e., $\int \Psi^{-1}_{\bm{n}'}(\bm{q}) \Psi_{\bm{n}}(\bm{q}) ~\mathrm{d} \bm{q} = \delta_{\bm{n}'\bm{n}}$.
Using the normalizing flow, 
the phonon coordinates $\bm{q}$ are connected to the quasi-phonon coordinates $\bm{\xi}$ through a coordinate transformation, 
also known as point transformation~\cite{eger1963Point, eger1964point, dewitt1952point}.
In our work, it can be understood as a reversible function $\bm{\xi}=f_{\bm{\theta}}(\bm{q})$,
where $f_{\bm{\theta}}$ is realized through neural networks~\cite{dinh2016density} with learnable parameters $\bm{\theta}$.

\subsection{B. Gradients of the variational free energy}

Under constant-volume conditions, 
the variational Helmholtz free energy is calculated through the expectation values of the phonon energy occupation probabilities $p_{\bm{n}}$ and phonon wave functions $\Psi_{\bm{n}}(\bm{q})$:
\begin{equation}
 F = \underset{\bm{n} \sim p_{\bm{n}}}
        {\mathbb{E}} \left[
 k_B T \ln p_{\bm{n}} + 
        \underset{\bm{q} \sim |\Psi_{\bm{n}}(\bm{q})|^2}
        {\mathbb{E}} 
        \left[
 E_{\bm{n}}^{\text{vib}}(\bm{q})
        \right]
    \right],
\label{eq:f}
\end{equation}
where the symbol ${\mathbb{E}}$ represents statistical expectation, which can be estimated through sampling, and the number of samples is called the batch size $B$~\cite{zhang2024neural, becca2017quantum}.
For instance, consider an observable $O(x)$, a function of a variable $x$ that follows a probability distribution $\pi(x)$.
Let $b$ denotes the index of each sample, where $b = 1, 2, \dots, B$, and $O(x_b)$ represents the value corresponding to the $b$-th sample. 
The unbiased estimation of the expectation can be approximated using these samples:
\begin{equation}\label{eq:observable}
\int  O(x) ~ \pi(x) ~\mathrm{d} x
=
\underset{x \sim \pi(x)}{\mathbb{E}} \left[ O(x) \right] 
=
 \frac{1}{B} \sum_{b=1}^B O(x_b).
\end{equation}
Returning to Eq.~\ref{eq:f}, the term $E_{\bm{n}}^{\text{vib}}(\bm{q}) = \Psi^{-1}_{\bm{n}}(\bm{q}) H_{\text{vib}} \Psi_{\bm{n}}(\bm{q})$ represents the local energy of the vibrational Hamiltonian.
It is determined by the energy level indices $\bm{n}$ and the phonon coordinates $\bm{q}$:
\begin{equation}
 E_{\bm{n}}^{\text{vib}}(\bm{q}) 
 = - \frac{1}{2} \sum_{k=1}^D 
    \left[
        \frac{\partial^2}{\partial q_k^2} \ln |\Psi_{\bm{n}}(\bm{q})|
        + \left(
            \frac{\partial}{\partial q_k} \ln |\Psi_{\bm{n}}(\bm{q})|
        \right)^2
    \right]
    + \left[
    \frac{1}{2} \sum_{k=1}^D  \omega_k^2 q_k^2
    + V_{\text{anh}}(\bm{q})
    \right]
    ,
    \label{eq:eloc}
\end{equation}
where the first part is kinetic energy and the second part is potential energy.
Since the wave functions are real-valued, 
we directly take the logarithm of their absolute values.
This technique helps prevent numerical issues during computation, 
as detailed in the Appendix of Ref.~\cite{zhang2024neural}.
Moreover,
the local energy's computational bottleneck lies in evaluating the second-order derivatives. 
To reduce the computational complexity, we adopt Hutchinson's stochastic trace estimator~\cite{hutchinson1989stochastic}. 
This trick improves the computational efficiency by an order of magnitude without compromising accuracy~\cite{xie2023mstar}.
The free energy $F$ can be minimized by optimizing the parameter $\bm{\mu}$ in the probabilities $p_{\bm{n}}$ and the parameter $\bm{\theta}$ in the wave functions $\Psi_{\bm{n}}(\bm{q})$.
The gradients of $F$ to these two parameters are given by~\cite{xie2022abinitio, xie2023mstar, zhang2024neural}:
\begin{equation}
\begin{aligned}
\nabla_{\bm{\mu}} F &= \
    \underset{\bm{n} \sim p_{\bm{n}}}{\mathbb{E}} \left[
    \left(
            \nabla_{\bm{\mu}} \ln p_{\bm{n}}
    \right) \cdot
    \left(
k_B T \ln p_{\bm{n}} + 
    \underset{\bm{q} \sim |\Psi_{\bm{n}}(\bm{q})|^2}{\mathbb{E}} \left[
 E_{\bm{n}}^{\text{vib}}(\bm{q})
    \right]
    \right)
\right], \\
\nabla_{\bm{\theta}} F &= 
    2 \underset{\bm{n} \sim p_{\bm{n}}}{\mathbb{E}} 
    \left[
      \underset{\bm{q} \sim |\Psi_{\bm{n}}(\bm{q})|^2}{\mathbb{E}} 
        \left[
            \left( \nabla_{\bm{\theta}} \ln |\Psi_{\bm{n}}(\bm{q})| \right) \cdot
            \left( E_{\bm{n}}^{\text{vib}}(\bm{q})  \right)
        \right]
        \right].
\end{aligned}
\end{equation}

\subsection{C. Observables and anharmonic zero-point energy}

During the training process of NCT, observables such as kinetic energy, potential energy, entropy, and Helmholtz free energy are computed at each iteration. 
This approach also enables the convenient computation of pressure and stress, which is crucial for structural relaxation in the isothermal-isobaric ensemble. 
This implementation utilizes automatic differentiation~\cite{jax2018github, jaxmd2020}.
The stress tensor $\bm{\sigma}$, which accounts for both nuclear quantum and anharmonic effects, can be directly derived from its definition~\cite{nielsen1985quantum,martin2020electronic}:
\begin{equation}
\sigma_{\alpha\beta} =
- \frac{1}{\Omega} 
\left.
\frac{\partial F}{\partial \varepsilon_{\alpha\beta}}
\right|_{\varepsilon=0}
= - \frac{1}{\Omega}
\left\langle 
\sum_i \frac{1}{2m_i} \nabla_{i\alpha} \nabla_{i\beta}
- 
\frac{1}{2}
\sum_{i\neq j}
\frac{(\bm{r}_{ij})_\alpha (\bm{r}_{ij})_\beta} {|\bm{r}_{ij}|}
\frac{\mathrm{d} V_{\text{el}}(|\bm{r}_{ij}|)} {\mathrm{d}|\bm{r}_{ij}|}
\right\rangle,
\end{equation}
where $\Omega$ is the volume of the system, 
and the strain $\varepsilon$ is defined as a transformation of each atomic coordinates
$r_{\alpha}=(\delta_{\alpha\beta} + \varepsilon_{\alpha\beta})r_{\beta}$.
The term ${\mathrm{d} V_{\text{el}}(|\bm{r}_{ij}|)}/ {\mathrm{d}|\bm{r}_{ij}|}$ can be interpreted as the force, and it can be conveniently derived from BOES.
Then, the pressure is determined from the trace of the stress tensor:
\begin{equation}
P = - \frac{1}{3} \sum_\alpha \sigma_{\alpha\alpha} 
= \frac{1}{3\Omega}
\left(
    2E_k -
    \frac{1}{2}
\left\langle 
\bm{r}_{ij} \frac{\mathrm{d} V_{\text{el}}(|\bm{r}_{ij}|)} {\mathrm{d}|\bm{r}_{ij}|}
\right\rangle
\right),
\end{equation}
where $E_k$ is the kinetic energy of the ions, which can be derived from the first part of Eq.~\ref{eq:eloc}.
In the study of lithium, we observe that the electronic contribution to pressure significantly outweighs the phononic contribution.
For example, under $70~\text{GPa}$, the phonon contribution to the pressure is less than $2~\text{GPa}$, with the remainder arising from electrons.


In a supercell containing $N$ atoms, which has $D=3N-3$ phonons.
The zero-point energy (ZPE) in the harmonic approximation is defined using the frequencies $\omega_k$ derived from the dynamical matrix:
$E_{\text{ZPE}, \text{har}} = \sum_{k=1}^D \omega_k/{2N}$.
The anharmonic phonon frequencies $w_k$ and anharmonic ZPE in NCT can be defined as follows.
The energy expectation for the energy level indexed by $\bm{n}=(n_1, n_2, \ldots, n_D)$ can be computed through sampling~\cite{zhang2024neural}:
\begin{equation}
    E_{\bm{n}} 
    = 
    \int \Psi^{-1}_{\bm{n}}(\bm{q}) H_{\text{vib}} \Psi_{\bm{n}}(\bm{q}) ~\mathrm{d} \bm{q}
    =
        \underset{\bm{q} \sim |\Psi_{\bm{n}}(\bm{q})|^2}
        {\mathbb{E}} 
        \left[
            E_{\bm{n}}^{\text{vib}}(\bm{q})
        \right],
\end{equation}
where $E_{\bm{n}}^{\text{vib}}(\bm{q})$ is the local energy defined in Eq.~\eqref{eq:eloc}.
In this context, the first excited single-phonon state of the $k$-th phonon is represented as
$(n_1, n_2, \ldots, n_D) = (0, 0, \ldots, 0)$, except for $n_k=1$, and the second excited state corresponds to $n_k = 2$.
The anharmonic frequency of each phonon is defined by subtracting the ground-state energy from its single-phonon excitation energy:
$
    w_k = E_{n_k=1} - E_{n_k=0},
$
where the reduced Planck constant is set to be $\hbar=1$ for simplicity.
The anharmonic ZPE is then computed as the half sum of these frequencies, normalized by the number of atoms:
\begin{equation}
    E_{\text{ZPE}, \text{anh}} = \frac{1}{N}\sum_{k=1}^D \frac{w_k}{2}.
\end{equation} 

\section{SII. Computational details and Additional results}

\subsection{A. Benchmarking on a one-dimensional anharmonic potential}

\begin{figure}[htbp]
	\centering
    \hspace{0mm}
	\includegraphics[width=0.65\textwidth]{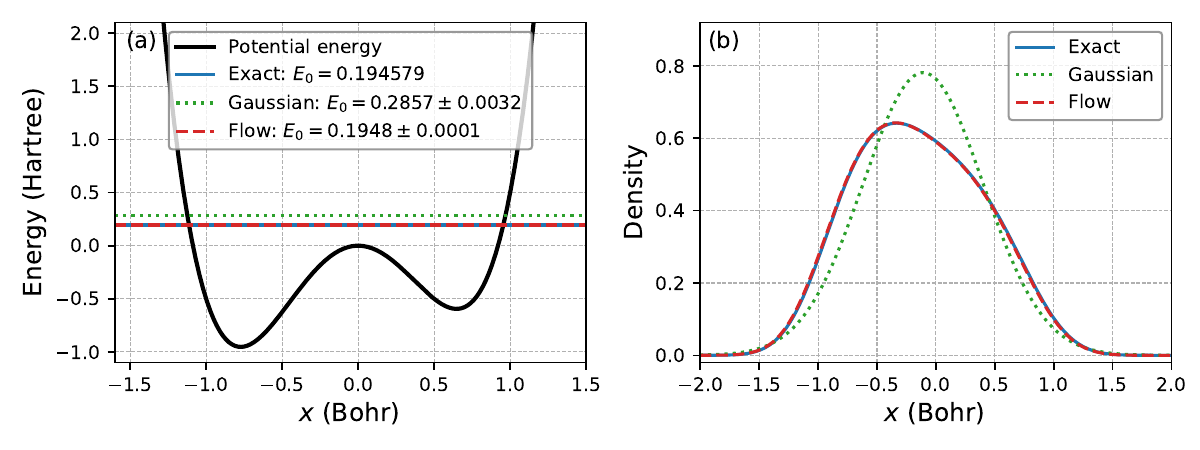}
	\caption{
    A one-dimensional particle in an anharmonic potential.
    (a) Potential energy surface (black curve) along with the corresponding ground state energies $E_0$ (horizontal lines). 
    The legend ``Exact" refers to the exact solution. 
    ``Gaussian" denotes the result obtained using a Gaussian-type wave function, which aligns with the assumptions in the stochastic self-consistent harmonic approximation (SSCHA).
    ``Flow" refers to the normalizing flow-based variational wave function implemented in neural canonical transformation (NCT).
    The numbers following the ``$\pm$" represent the statistical errors arising from sampling.
    (b) The quantum distribution of the ground state $|\Psi(x)^2|$.
    }
	\label{fig:s0}
\end{figure}

To explicitly contrast NCT's advantage over the stochastic self-consistent harmonic approximation (SSCHA), we carry out a benchmark study on the one-dimensional double-well potential considered in the SSCHA review~\cite{monacelli2021sscha}, given by $V(x) = 3 x^4 + \frac{1}{2} x^3 - 3 x^2$.
All quantities are expressed in atomic units: Bohr (length), Hartree (energy), and the particle mass is set to $1$ amu.
The exact solution is obtained using the finite difference method. 
A Gaussian-type wave function ansatz, $\Psi(x)=e^{-\frac{1}{2} w x^2}$, undergoes variational optimization of $w$ to replicate results from the SSCHA. 
In NCT, a single-particle variational wave function $\Psi(x) = e^{-\frac{1}{2} \omega f_{\bm{\theta}}(x)^2}\sqrt{\left| \partial f_{\bm{\theta}}(x) / \partial x \right|}$ is employed, where the bijection $f_{\bm{\theta}}(x)$ is constructed using an $8$-layer residual network~\cite{he2016deep}.

The results are shown in FIG.~\ref{fig:s0}.
In the first subplot, FIG.~\ref{fig:s0}~(a), 
it is evident that with the help of normalizing flow, the NCT method yields a result that is only slightly higher than the exact solution, with an energy difference on the order of $10^{-4}~\text{Ha}$.
In contrast, the Gaussian-type wave function exhibits significantly larger errors when applied to the strongly anharmonic potential~\cite{monacelli2021sscha}, overestimating the exact energy by approximately $0.09~\text{Ha}$.
This discrepancy is even more pronounced in the wave function density, as shown in FIG.~\ref{fig:s0}~(b). 
The result obtained using normalizing flow closely matches the exact solution, whereas the Gaussian approximation deviates substantially, yielding a symmetric curve that fails to capture the asymmetric features of the anharmonic potential~\cite{monacelli2021sscha}.

\subsection{B. Born-Oppenheimer energy surface and computational costs}

In this work, the BOES is realized through the deep potential (DP) models~\cite{wang2018deepmd, zeng2023deepmd, wang2023data},
which are trained on datasets generated through density functional theory (DFT) calculations.
The DFT calculations were performed using the VASP (Vienna ab initio simulation package) code~\cite{vaspref}, employing the projector augmented wave (PAW) method with the Perdew-Burke-Ernzerhof (PBE) functional~\cite{perdew1996generalized}.
For high-pressure calculations of \textit{cI16} and \textit{oC88}, we utilized the same DP model (Li-DP-Hyb2) as detailed in the Supplemental Material of Ref.~\cite{wang2023data}.
At low pressures, another DP model (Li-DP-Hyb3) was trained with additional data for \textit{bcc} and \textit{fcc} structures.

\begin{table}[htbp]
    \centering
    \caption{Computational costs for neural canonical transformations.
    }
    \label{tab:costs}
    \setlength{\tabcolsep}{5pt}
    \begin{tabular}{C{2.2cm}C{2.0cm}C{2.0cm}C{2.0cm}C{6.5cm}}
        \addlinespace[3pt]
        \hline\hline
        \addlinespace[3pt]
        \textbf{structure} & 
        \makecell{\textbf{atoms}} & 
        \makecell{\textbf{batch size}} & 
        \makecell{\textbf{epochs}} & 
        \makecell{\textbf{GPU time (hours)}} \\
        \addlinespace[3pt]
        \hline
        \addlinespace[3pt]
        \textit{bcc}  & 250 & 400 & 15000 & 2*A100: 18 h or 4*V100: 26 h \\
        \textit{fcc}  & 256 & 400 & 15000 & 2*A100: 18 h or 4*V100: 26 h \\
        \textit{cI16} & 432 & 256 & 15000 & 4*A100: 22 h or 8*V100: 29 h \\
        \textit{oC88} & 352 & 256 & 67000 & 2*A800: 120 h or 4*V100: 140 h \\
        \addlinespace[3pt]
        \hline\hline
    \end{tabular}
\end{table}

In the NCT calculations presented in this work, the probability distribution of phonon energy levels $p_{\bm{n}}$ is modeled using the product spectrum ansatz, with energy levels truncated at $K=20$. 
The function $f_{\bm{\theta}}$ in the variational wave functions is implemented using a real-valued non-volume preserving network~\cite{dinh2016density}, consisting of $6$ coupling layers, each comprising two multilayer perceptrons with $128$ hidden units. 
For the orthorhombic crystals, \textit{oC88}, structure relaxation is performed every $4000$ iterations to optimize the lattice constant $\bm{a}$.
For \textit{bcc} and \textit{fcc} structures, relaxation occurs every $50$ iterations. 
The NCT method demonstrates exceptional scalability for large-scale parallel computations. 
All calculations were conducted on multiple NVIDIA A800-80G or V100-32G GPUs (graphics processing units). 
TABLE~\ref{tab:costs} summarizes the computation time for each data point, 
where a single data point corresponds to a specific combination of structure, pressure, and temperature. 
Additional details regarding parameter settings can be found in our code repository~\cite{mycode}.

There are several sources of uncertainty in our calculations, which we aim to quantify or eliminate below. 
First, the error bars on all figures represent the statistical errors from Eq.~\ref{eq:observable}. 
A large number of samples ($80,000$ for \textit{bcc} and \textit{fcc}, and $51,200$ for \textit{cI16} and \textit{oC88}) were used to calculate the thermodynamic averages, thereby reducing statistical errors.
The typical magnitudes of the statistical errors are: pressure of $0.001 ~\text{GPa}$, energy of $0.01~\text{meV/atom}$, and entropy of $0.0002~k_B/\text{atom}$. 
Second, additional systematic errors arise from the density functional and the variational approximation.
In the study of \textit{bcc} and \textit{fcc}, the errors introduced by the Li-DP-Hyb3 model range from $0.2\sim0.5~\text{meV/atom}$, but these are systematic. Since our classical molecular dynamics (MD) comparison studies use the same DP model, and the contribution of anharmonic entropy to the free energy is more significant than the DP model error, these uncertainties do not change our conclusions.
In the analysis of the \textit{cI16}-\textit{oC88} transition, the errors introduced by the Li-DP-Hyb2 model are less than $1~\text{meV/atom}$, which is smaller than the effects of nuclear quantum effects and various DFT functionals. We further verified this through direct DFT calculations, which confirm that these errors do not alter our conclusions.
Lastly, in order to decrease the finite-size effects, we used nearly equal supercells for \textit{bcc} ($250$ atoms) and \textit{fcc} ($256$ atoms), consistent with classical MD simulations.
For the \textit{oC88} and \textit{cI16} structures, the largest feasible supercells are used, employing $352$ atoms for \textit{oC88} and $432$ atoms for \textit{cI16}, in line with those used in MD.

\subsection{C. Phase diagrams via neural canonical transformations}

\begin{figure}[htbp]
	\centering
    \hspace{0mm}
	\includegraphics[width=0.75\textwidth]{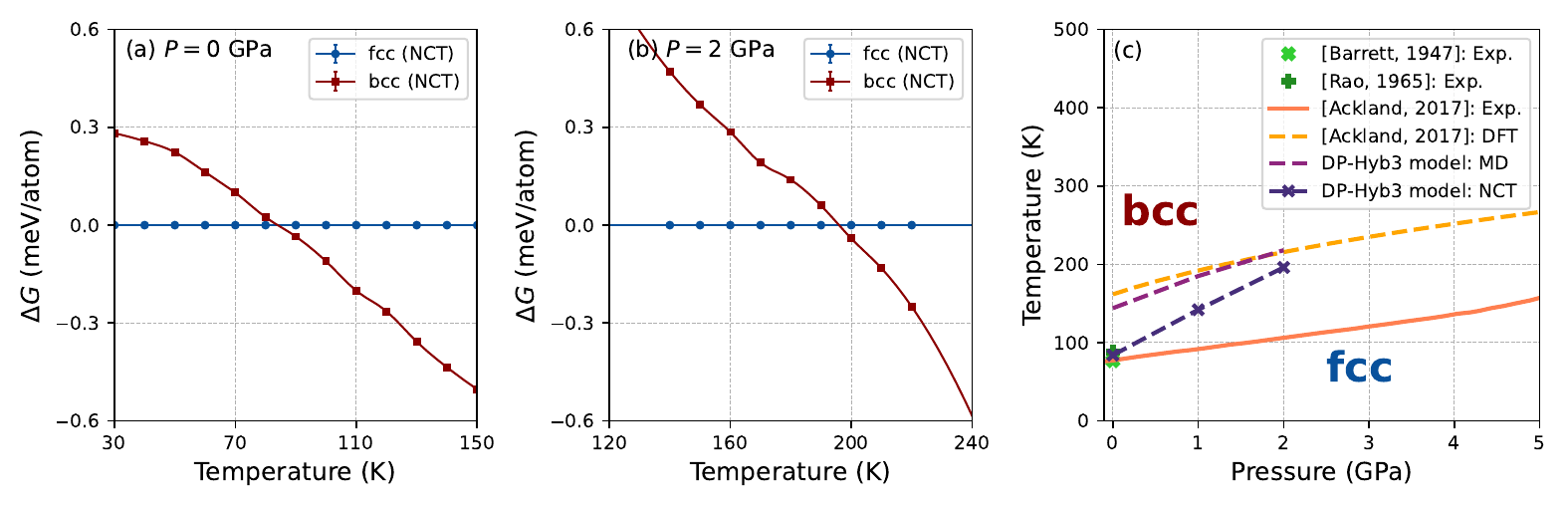}
	\caption{
    (a) Gibbs free energy difference between the \textit{fcc} and \textit{bcc} structures at $P = 0~\text{GPa}$, calculated through neural canonical transformations (NCT). 
    (b) At $P = 2~\text{GPa}$.
    (c) Phase diagram for \textit{fcc}-\textit{bcc} transitions. 
    The experimental transition points at $0~\text{GPa}$ are taken from Refs.~\cite{barrett1947a, rao1967phase}.
    The experimental transition line and density functional theory (DFT) results are sourced from Ref.~\cite{ackland2017Quantum}. 
    Both classical molecular dynamics (MD) and NCT calculations utilize the same deep potential model (Li-DP-Hyb3).
    }
	\label{fig:s1}
\end{figure}

We utilized the NCT method to calculate the Gibbs free energy for \textit{fcc} and \textit{bcc} structures, as depicted in FIG.~\ref{fig:s1}. 
NCT predicts transition temperatures of $84$, $142$, and $196~\text{K}$ at $0$, $1$, and $2~\text{GPa}$, which are consistently lower than those obtained from thermodynamic integration through classical molecular dynamics (MD) simulations on the same energy surface. 
The latter yield transition temperatures of $144$, $185$, and $218~\text{K}$.
This discrepancy is primarily attributed to the quantum effects of nuclei.
Notably, at $0~\text{GPa}$, the MD-predicted phase boundary of $144~\text{K}$ aligns closely with the DFT value reported in Ref.~\cite{ackland2017Quantum}. 
However, NCT significantly lowers the phase boundary to $84~\text{K}$, which agrees more closely with experimental values of $77$ and $88~\text{K}$ in Refs.~\cite{barrett1947a, rao1967phase}.
Compared to the MD results, NCT provides corrections in the right direction, underscoring the importance of incorporating nuclear effects.

\begin{figure}[htbp]
	\centering
    \hspace{0mm}
	\includegraphics[width=0.75\textwidth]{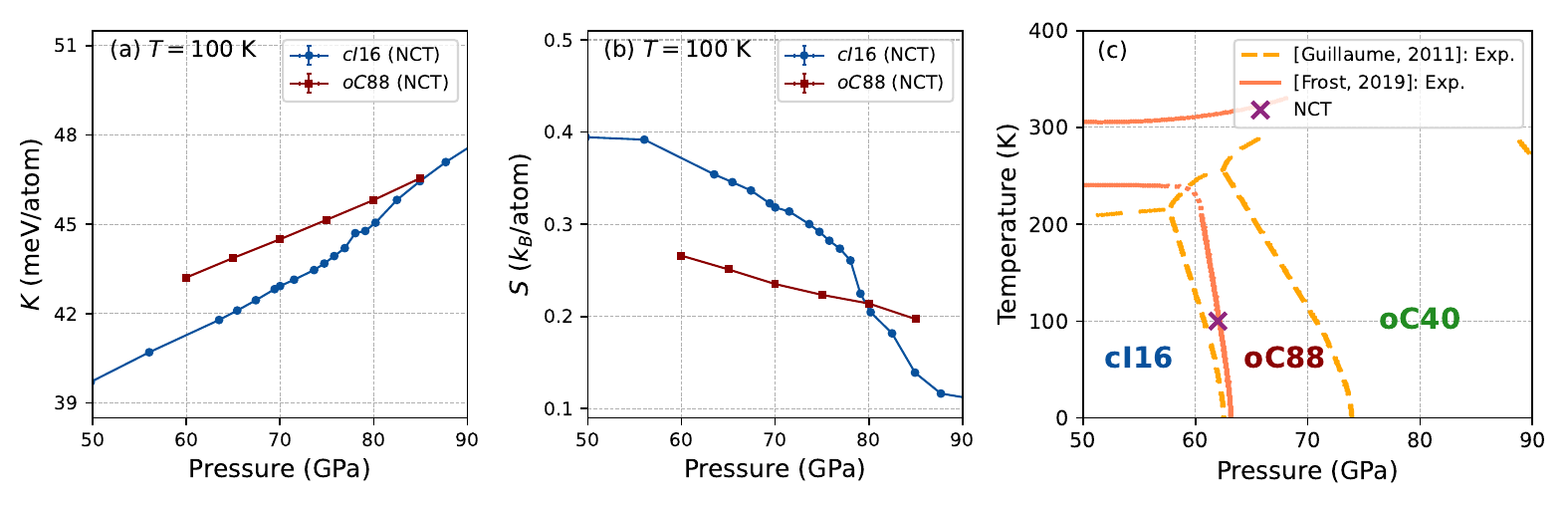}
	\caption{
    (a) Kinetic energy of \textit{cI16} and \textit{oC88} structures at $T=100~\text{K}$, calculated through neural canonical transformations (NCT). 
    (b) Ironic entropy.
    (c) Phase diagram. 
    The experimental transition lines are taken from Refs.~\cite{guillaume2011cold,frost2019high}.
    The NCT result has been corrected.
    }
	\label{fig:s2}
\end{figure}

In the main text, we discussed calculations for the \textit{cI16} and \textit{oC88} structures using the NCT method.
The results for ionic kinetic energy and entropy are presented in FIGs.~\ref{fig:s2}~(a) and (b), respectively. 
Within the pressure range of $60$ to $70~\text{GPa}$, the \textit{cI16} structure consistently exhibits lower kinetic energy and higher entropy.
However, this outcome contradicts the hypothesis proposed in Ref.~\cite{gorelli2012lattice}, which suggested that \textit{oC88} might exhibit lower ZPE and higher entropy than the competing structures.
At pressures above $75~\text{GPa}$, both \textit{cI16} and \textit{oC88} become unstable, and a more stable semiconducting \textit{oC40} structure emerges.
The accuracy of the DP model also declines significantly in this range, making the results for \textit{cI16} and \textit{oC88} less significant. 
FIG.~\ref{fig:s2}~(c) shows the high-pressure phase diagram, where the NCT-derived \textit{cI16}-\textit{oC88} phase boundary is corrected using hybrid functionals. 
The transition point agrees well with experimental observations~\cite{guillaume2011cold,frost2019high}.

\subsection{D. High-accuracy electronic calculations at high-pressure}

\begin{table}[htbp]
    \centering
    \caption{Potential energy, enthalpy, and Gibbs free energy differences for the \textit{cI16} and \textit{oC88} structures, calculated via various electronic structure methods. 
    The structures are obtained via NCT at $T=100~\text{K}$ and $P=70~\text{GPa}$.
    All values in the table are given in $\text{meV}/\text{atom}$, and take \textit{cI16} as the reference.
    (1) Potential energies are calculated at equilibrium positions: $V = V_{\text{el}}(\bm{R_0})$.
    (2) Enthalpies are determined as $H = P\Omega + V$, where $P=70~\text{GPa}$ and $\Omega$ is the volume per atom.
    (3) Gibbs free energy differences are corrected based on neural canonical transformations (NCT) presented in the main text: $\Delta G = \Delta G_\text{NCT} + \Delta V$,
    where we set $\Delta G_\text{NCT} = 3.96~\text{meV}/\text{atom}$ for PBE.
    }
    \label{tab:dft}
    \setlength{\tabcolsep}{5pt}
    \begin{tabular}{C{0.8cm}L{1.8cm}C{2.2cm}C{2.2cm}C{2.2cm}C{2.2cm}C{2.2cm}}
        \addlinespace[3pt]
        \hline\hline
        \addlinespace[3pt]
        \multicolumn{2}{c}{\textbf{Method}}           & \textbf{DP~(VASP)}         & \textbf{ABACUS}     & \textbf{ABACUS}   & \textbf{FHI-aims} & \textbf{FHI-aims} \\
        \multicolumn{2}{c}{Functional}       & PBE        & PBE        & HSE06      & PBE      & HSE06    \\
        \multicolumn{2}{c}{Basis}            & Plane Wave & DZP        & DZP        & intermediate        & intermediate        \\
        \multicolumn{2}{c}{Pseudopotential} & PAW & Dojo-NC-FR & Dojo-NC-FR & -        & -        \\
        \addlinespace[3pt]
        \hline
        \addlinespace[3pt]
        $\Delta V$ &\textit{oC88} - \textit{cI16} 
        & \textcolor{red}{$+7.06$}  
        & \textcolor{red}{$+7.82$}  & \textcolor{red}{$+1.65$} 
        & \textcolor{red}{$+8.82$}  & \textcolor{red}{$+1.15$} \\
        $\Delta H$ &\textit{oC88} - \textit{cI16} 
        & \textcolor{red}{$+1.68$}  
        & \textcolor{red}{$+2.44$}  & \textcolor{green}{$-3.73$} 
        & \textcolor{red}{$+3.44$} & \textcolor{green}{$-4.23$} \\
        $\Delta G$ &\textit{oC88} - \textit{cI16}
        & \textcolor{red}{$+3.96$}  
        & \textcolor{red}{$+3.96$}  & \textcolor{green}{$-2.21$}
        & \textcolor{red}{$+3.96$}  & \textcolor{green}{$-3.71$} \\
        \addlinespace[3pt]
        \hline\hline
    \end{tabular}
\end{table}

Further DFT calculations were conducted using the ABACUS software package~\cite{abacuspackage} (TABLE.~\ref{tab:dft}, columns 3 and 4), which utilizes the Dojo norm-conserving fully-relativistic pseudopotentials (Dojo-NC-FR) generated with the ONCVPSP code (version 3.3.0)~\cite{abacusref2,abacusref1}. 
A double-zeta polarized (DZP) basis set was employed for the Perdew-Burke-Ernzerhof (PBE) and Heyd-Scuseria-Ernzerhof (HSE06) hybrid functional calculations.
Additional calculations were performed using the FHI-aims all-electron electronic structure code~\cite{aims} (TABLE.~\ref{tab:dft}, columns 5 and 6), which leverages numerically tabulated atom-centered basis functions for highly accurate and efficient quantum mechanical simulations~\cite{aimsref1}. 
An intermediate basis set was chosen to balance computational cost and accuracy.
The HSE06 hybrid functional was applied to capture the exchange-correlation effects, utilizing the standard mixing scheme of $1/4$ Hartree-Fock and $3/4$ Kohn-Sham components~\cite{aimsref2, martin2020electronic}.

We performed single-point calculations on the structures optimized via NCT at $100~\text{K}$ and $70~\text{GPa}$, with the results shown in TABLE~\ref{tab:dft}.
For the \textit{cI16} structure, a $15 \times 15 \times 15$ $k$-point grid was utilized, while an $8 \times 8 \times 8$ grid for \textit{oC88}. All input files for these calculations are accessible in the code repository~\cite{mycode}.
The second column presents the results obtained using the DP model, trained on data generated by VASP with the PAW method under the PBE functional.
The third column presents the results from ABACUS using a DZP basis under PBE, which aligns with those of the DP model, confirming that the DP model's accuracy does not influence the conclusions.
In this case, the potential energy at the equilibrium position of \textit{oC88} is $7.82~\text{meV}/\text{atom}$ higher than that of \textit{cI16}.
The fourth column presents the results under the hybrid HSE06 functional, where the energy difference between the two structures is reduced, with \textit{oC88} being only $1.65~\text{meV}/\text{atom}$ higher than \textit{cI16}.
This indicates that HSE corrects the BOES by $6.17~\text{meV}$ compared to PBE.
Similar results were obtained through calculations performed with FHI-aims,
where HSE yielded a correction of $7.67~\text{meV}/\text{atom}$ compared to PBE.

To further investigate the stability of these two structures, we estimated their enthalpies, incorporating corrections based on single-point electronic structure calculations.
It is evident that when a more accurate functional is employed, \textit{oC88} becomes more stable than \textit{cI16}. 
This suggests its stabilization originates from electronic structure effects rather than finite-temperature or nuclear quantum effects.
In addition, we applied corrections to the Gibbs free energy obtained from the NCT calculations.
Interestingly, quantum and thermal effects slightly reduce the stability of the \textit{oC88} structure, contradicting the hypotheses proposed in previous studies~\cite{marques2011crystal, gorelli2012lattice, wang2023data}.

\end{widetext}

\end{document}